\documentclass[aip,graphicx]{revtex4-1}
\usepackage[cp1251]{inputenc}
\usepackage[T2A]{fontenc}
\usepackage[english]{babel}
\usepackage{amssymb,latexsym,amsmath,amscd}
\usepackage{graphicx,color,framed}
\begin{document}
%\selectlanguage{russian}
\selectlanguage{english}

\title{Polymer chain collapse induced by many-body dipole correlations}

\author{\firstname{Yu.~A.} \surname{Budkov}}
\email[]{ybudkov@hse.ru}
\email[]{urabudkov@rambler.ru}
%\homepage[]{Your web page}
%\thanks{}
%\altaffiliation{}
\affiliation{National Research University Higher School of Economics, Department of Applied Mathematics, Moscow, Russia}
\author{\firstname{N.~N.} \surname{Kalikin}}
%\email[]{bancocker@mail.ru}
%\homepage[]{}
%\thanks{}
%\altaffiliation{}
\affiliation{Ivanovo State University, Department of Physics, Ivanovo, Russia}
\author{\firstname{ A.~L.} \surname{Kolesnikov}}
%\email[]{bancocker@mail.ru}
%\homepage[]{}
%\thanks{}
%\altaffiliation{}
\affiliation{Institut f\"{u}r Nichtklassische Chemie e.V., Universit\"{a}t Leipzig, Leipzig, Germany}

\begin{abstract}
We present a simple analytical theory of flexible polymer chain dissolved in a good solvent, carrying permanent freely oriented dipoles on the monomers. We take into account the dipole correlations within the random phase approximation (RPA), as well as a dielectric heterogeneity in the internal polymer volume relative to the bulk solution. We demonstrate that the dipole correlations of monomers can be taken into account as pairwise ones only when the polymer chain is in a coil conformation. In this case the dipole correlations manifest themselves through the Keesom interactions of the permanent dipoles. On the other hand, the dielectric heterogeneity effect  (dielectric mismatch effect) leads to effective interaction between the monomers of the polymeric coil. Both of these effects can be taken into account by the renormalizing the second virial coefficient of the volume interactions monomer-monomer. We establish that in the case when the solvent dielectric permittivity exceeds the dielectric permittivity of the polymeric material, the dielectric mismatch effect competes with the dipole attractive interactions, leading to polymer coil expansion. In the opposite case, both the dielectric mismatch effect and the dipole attractive interaction lead to the polymer coil collapse. We analyse the coil-globule transition caused by the dipole correlations of monomers within the many-body theory. We demonstrate that accounting for the dipole correlations higher than pairwise ones smooths this pure electrostatics driven coil-globule transition of the polymer chain.
\end{abstract}

\maketitle
\section{Introduction}
The coil-globule (CG) transition in dilute polymer solutions is a very important phenomenon for various technological advances, ranging from polymer functionalizations (such as polymerization, plasticization, dyeing, etc.) in chemical
industry \cite{Tomasko,Kazarian,Kiran2016} to encapsulation of drug compounds into polymer globules and their subsequent targeted delivery in pharmaceutical applications \cite{DrugDeliveryReview1,DrugDeliveryReview2,DrugDeliveryReview3}.
Therefore, the theoretical importance of the CG transition
has attracted great attention of many researchers during the last few decades. Both theorists and experimentalists in soft matter physics have taken great efforts to develop CG transition theory.

The existing theoretical models made a large contribution to an understanding of this phenomenon \cite{deGennes_collaps,Grosberg, Birshtein,Moore,Sanchez,Lifshitz,Lifshitz1,Muthukumar,Dua,Vilgis2005,Simmons2013,Muzdalo,Matsuyama1990,Tamm,
Budkov1,Budkov2,Budkov3,Budkov4}. The classical theoretical models \cite{deGennes_collaps,Grosberg,Birshtein,Moore,Sanchez,Lifshitz}
of the CG transition are based on the idea that decreasing of the solution temperature below a certain threshold value (theta-temperature)
leads to the domination of the attractive interactions between the monomers and, thus, to the collapse of the polymer coil.
Such a simple idea allows one to rationalize the conformational behavior of real synthetic polymer chains in the solvent media.

It is well known from the experiments (see, for instance, \cite{Melnikov1999,Borkovec2004}) and MD simulations \cite{Brilliantov2016,Gavrilov2016,Netz_2003,Netz_2003_2}, that in dilute polyelectrolyte solutions in the
regime of good solvent, the CG transition of the flexible polyelectrolyte chain can take place. This unconventional
CG transition of the polyelectrolyte chain is accompanied by counterion condensation \cite{Manning1978}.
Therefore, this CG transition is purely electrostatic in nature. Due to the fact that this electrostatic conformational transition takes place in the regime of good solvent, i.e., when the polyelectrolyte chains are well soluble, we cannot use the mentioned above classical models of the CG transition and, consequently, the concept of theta-temperature to describe it theoretically.

Two possible mechanisms of these electrostatic CG transitions have been so far proposed. The first mechanism (see original work \cite{Brilliantov1998}, where the {\sl counterion-fluctuation theory} was formulated) is based on the idea that the polymer chain collapse is caused by counterion electrostatic correlations \cite{Levin2002}. More specifically, when the electrostatic interactions become quite strong, the counterions prefer to adsorb onto the polymer surface, neutralizing the macromolecule charge. In this mechanism the counterions are not bounded strongly with the monomers, but can move freely along the polymer backbone (in this case the terms
'delocalized binding' or 'territorially bound' counterions are usually used \cite{Manning1978,Brilliantov1993}). Despite the full neutralizing of the macromolecule charge due to the counterion condensation \cite{Manning1978}, the thermal fluctuations of the charge density near its zero value are unavoidable. These charge density fluctuations, in turn, lead to cooperative mutual attraction of monomers \cite{Liverpool1999} (so-called Kirkwood-Shumaker interaction \cite{Kirkwood1952,Adzic2014}), causing the CG transition. It is worth noting that such correlation attraction of like-charged particles can cause phase separations in colloid and polyelectrolyte solutions (see, for instance, \cite{Brilliantov1993,Budkov_colloid,Budkov_polyel,Zhang2016,Shen2017}). Due to the fact that this CG transition takes place at rather high Coulomb strength (which is determined
as the characteristic electrostatic energy expressed in units of thermal energy $k_{B}T$  \cite{Muthu_collapse}) the electrostatic effects can be taken into account by going beyond the classical Debye-H\"{u}ckel (DH) theory. A successful
attempt to describe this electrostatic CG transition beyond the DH theory framework was taken in ref. \cite{Brilliantov1998} In this work the counterions that immersed to the background of the polymer chain charge were considered within the one component plasma (OCP) model, using the precise relation for the OCP electrostatic free energy \cite{BrilliantovOCP}. It is worth noting that accounting for the electrostatic correlations of counterions within the OCP model allowed researchers to rationalize the relations for the radius of gyration of the polyelectrolyte chain as a function of the Coulomb strength obtained from the MD simulations \cite{Brilliantov2016}. It should be noted that the electrostatic collapse of a highly charged polyelectrolyte chain in the regime of poor solvent has also been recently investigated theoretically, as well as by the MD simulation \cite{Brilliantov2017}. It should be noted that a theory of the polyelectrolyte collapse, where the counterion electrostatic correlations were taken into account at the DH theory level, was formulated by Kundagrami et al. in work \cite{Muthu_collapse}.

The second mechanism is based on the assumption that this CG transition takes place due to the attractive interaction of the thermally fluctuating dipoles, appearing along the polymer backbone due to the counterion condensation (the case of 'site bound' counterions \cite{Manning1978,Brilliantov1993}). This mechanism was first proposed in reference \cite{Schiessel1998} and later discussed in details in references \cite{Cherstvy2010,Dua2014}. However, within all mentioned above theories dipole
correlations were considered as pairwise. Indeed, in the works \cite{Schiessel1998,Dua2014} the influence of the dipole correlations on the polymer chain conformation was accounted for by renormalizing the second virial coefficient attributed to the volume interactions between monomers. In references \cite{Cherstvy2010,Kumar_2009} the dipole correlations were taken into account using the Keesom pair potential. As it has been recently showed in ref. \cite{Budkov6}, dipole correlations of monomers can be considered pairwise only when the polymer chain is in coil state. However, when the polymer chain is in the globular conformation, electrostatic dipole correlations must be taken into account at the many-body level. Therefore, the many-body electrostatic correlations might play a crucial role in the 'dipole' mechanism, as well as in the above discussed 'Coulombic' mechanism of the polyelectrolyte chain collapse.

Moreover, within all of these theoretical models \cite{Schiessel1998,Cherstvy2010,Kumar_2009,Dua2014,Budkov6} stated that the dielectric permittivity near the polymer backbone is the same that in the bulk solution. However, the dielectric heterogeneity near the polymer backbone relative to the bulk solution should be important in both of the above-mentioned mechanisms of the polyelectrolyte chain collapse.

Physically, in real polyelectrolyte solutions both the Coulombic and dipole scenarios of the CG transition could be realized. It depends mostly on the chemical specifics of the monomers and counterions. Thus, both of these mechanisms should be thoroughly analyzed from the first principles of statistical mechanics. However, to the best of our knowledge, analysis of the dipole mechanism of the CG transition in the dilute solutions of the electrically neutral polar polymers, regarding the many-body dipole correlations with an account for the dielectric heterogeneity has not been reported in the literature till now.

On the other hand, statistical physics of dielectric polar polymers remains one of the most undeveloped areas of polymer physics. Indeed, only several
theoretical works have been so far published, discussing thermodynamic and structural properties of dielectric polymers in the bulk solution without
\cite{Podgornik_2004,Kumar_2009,Dean_2012,Kumar_2014,Lu2015,Delaney2016,Mahalik2016} and with \cite{Gurovich1,Gurovich2,Budkov5,Budkov6} an electric field application.
In ref. \cite{Podgornik_2004} Podgornik studied within the Feynman path integrals formalism the behavior of the electrostatic persistence length of the semi-flexible polymer chain whose monomers interact through a screened dipolar interaction potential. In ref. \cite{Kumar_2009} Kumar et al. within the Edwards-Singh method calculated the mean-square radius of gyration of polyzwitterionic molecules in aqueous solutions depending on the different physico-chemical parameters, such as the chain length, electrostatic interaction strength, added salt concentration, dipole moment, and degree of ionization of the zwitterionic monomers. In ref. \cite{Dean_2012} the polarizing many-body correlations at the level of random phase approximation (RPA) were taken into account. Thereby, it was shown that the latter lead to ordering of the semi-flexible anisotropic polymer chains in the solution. In ref. \cite{Kumar_2014} Kumar et al. showed by means of the field-theoretic formalism that interactions between the monomeric dipoles in polymer blends lead to a considerable enhancement of the phase segregation. Lu et al. calculated within the field-theoretic formalism the interaction potential between two rigid polymers polarizable along their backbone depending on their mutual orientation \cite{Lu2015}. In ref. \cite{Delaney2016} the authors formulated a statistical field theory of the dielectric soft matter. In work \cite{Mahalik2016} the phase behavior of the polar polymer brushes depending on the dipole interaction strength was theoretically investigated. Works
\cite{Gurovich1,Gurovich2} theoretically studied the microphase separation in the co-polymer melts under the external electric field. Recent works have investigated the conformational behavior of the polarizable flexible polymer chain under the external electric field within the pure mean-field theory \cite{Budkov5} and the theory accounting for the many-body dipole correlations of monomers \cite{Budkov6}. It has been shown that in both theories regardless of the polymer chain conformation (coil or globule) increasing the electric field results in the polymer chain expansion (electrostriction). It has also been shown that the quite strong electric field in the regime of poor solvent can induce the globule-coil transition of the polarizable polymer chain due to the electrostriction effect.

In this paper we present a simple analytical self-consistent field theory of flexible polymer chain carrying the permanent freely oriented dipoles on the polymer backbone dissolved in a good solvent. We take into account the many-body dipole correlations within RPA, as well as the effect of dielectric heterogeneity near the polymer backbone relative to the bulk solution. We provide an analysis of the conformational behavior of the polar electrically neutral flexible polymer chain with an account for the dipole correlations of monomers at the many-body level. In our previous paper \cite{Budkov6} we mentioned shortly on the polymer chain collapse caused by the many-body dipole correlations. However, as already pointed out above, the main focus of our previous study was related to the globule-coil transition under the external electric field due to the electrostriction effect. Moreover, the details of the electrostatic free energy derivation were omitted. In the present study we focus on the CG transition induced by the many-body dipole correlations of monomers and provide a derivation of the electrostatic free energy of a flexible polymer chain with the freely oriented dipole moments on the monomers. In addition, we first study the influence of the dielectric mismatch effect on this pure electrostatic polymer chain collapse.

The paper is organized as follows: the theoretical background is presented in section 2, some analytical evaluations -- in section 3,  numerical results and discussion -- in section 4, and concluding remarks -- in section 5. At the end of the paper we placed the Appendix with some supplementary mathematical details.

\section{Theory}
We consider a dielectric flexible polymer chain with the degree of polymerization $N$ dissolved in a dielectric solvent with the permittivity $\varepsilon_{s}$. Let each monomer segment of the polymer chain carry the permanent freely oriented dipole moment $\bold{p}_{j}$ ($j=1,..,N$) (see Fig. 1). The latter may be realized for a weak polyelectrolyte chain in the regime of counterion condensation, when the counterions and monomers form the site bound ion pairs \cite{Manning1978,Brilliantov1993} or for the polyzwitterionic macromolecules \cite{Kumar_2009}. For the sake of simplicity in this study we will neglect the polarizability effect attributed to the fluctuations of the absolute value of the dipole moments. As one can show, accounting for this effect will not change the final outcomes. To study the conformational behavior of the polymer chain, we formulate a simple Flory-type \cite{Flory_book} self-consistent field theory, considering the radius of gyration $R_{g}$ as a single order parameter. Therefore, we assume that the polymer chain occupies the volume which can be estimated as the volume of gyration $V_{g}=4\pi R_{g}^3/3$.

\begin{figure}
\centerline{\includegraphics[scale = 0.5]{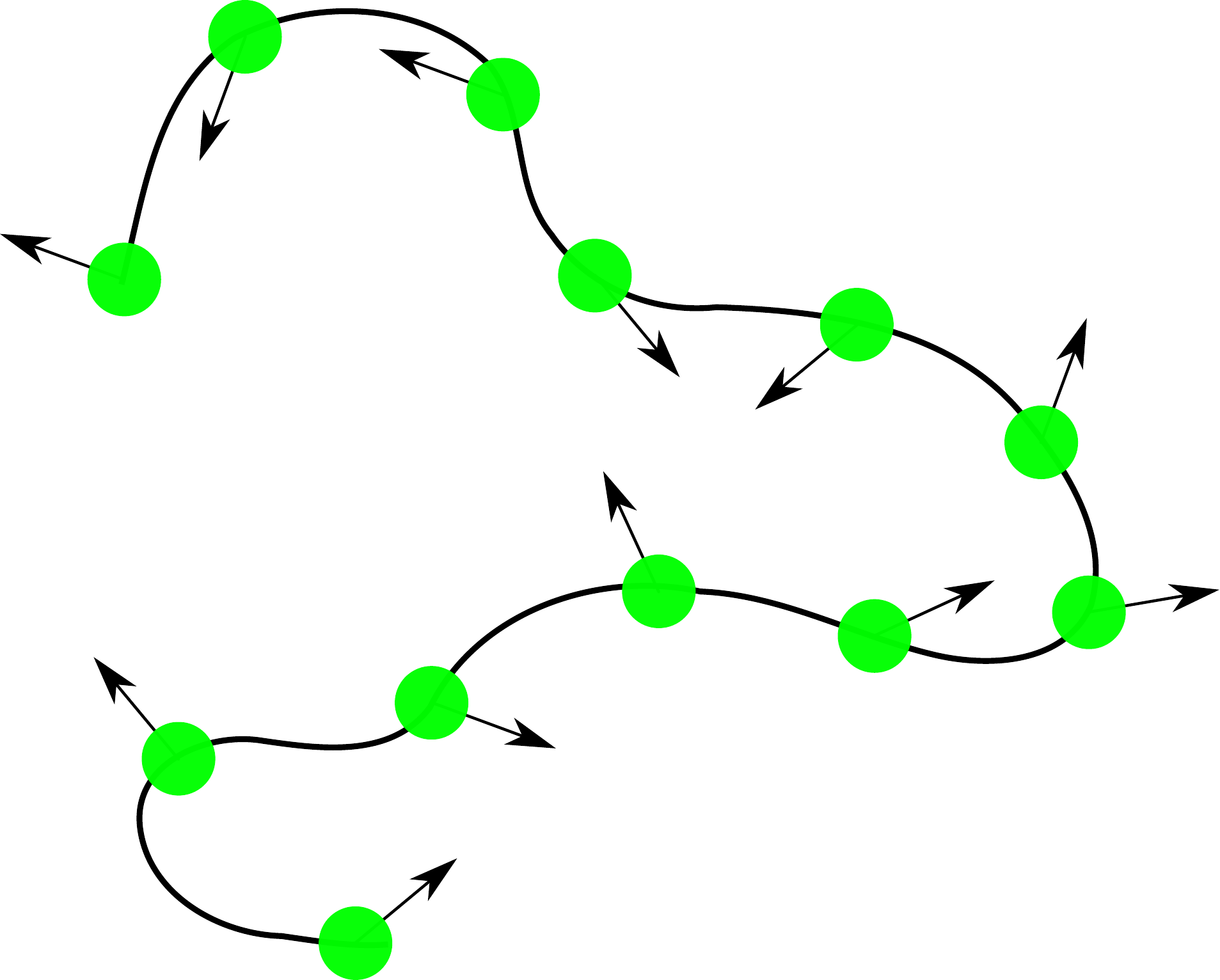}}
\caption{Illustration of the flexible polymer chain with permanent freely oriented dipole moments distributed along its backbone.}
\label{fig.1}
\end{figure}

The polymer chain total free energy can be written as a sum of three terms
\begin{equation}
\label{eq:Ftot}
F(R_{g})=F_{conf}(R_{g})+F_{vol}(R_{g})+F_{el}(R_{g}),
\end{equation}
where $F_{conf}(R_{g})$ is the conformation free energy of the ideal polymer chain which can be calculated by the following interpolation formula
\cite{Fixman,Grosberg,Budkov2016}
\begin{equation}
\label{eq:conf}
F_{conf}(R_{g})=\frac{9}{4}k_{B}T\left(\alpha^{2}+\alpha^{-2}\right),
\end{equation}
where $\alpha=R_{g}/R_{0g}$ is the expansion factor, $R_{0g}^2=Nb^2/6$ is the mean-square radius of gyration of the Gaussian polymer chain, $b$ is
the Kuhn length, $k_{B}$ is the Boltzmann constant, $T$ is the temperature. To take into account the volume interactions of monomers, we use the Flory-Huggins (FH) relation for excess free energy of the polymer chain in the solvent \cite{Matsuyama1990,Flory_book,Khokhlov_book}
\begin{equation}
\label{eq:Fvol}
F_{vol}(R_{g})=\frac{V_{g}k_{B}T}{v}\left((1-\phi)\ln{(1-\phi)+\phi-\chi \phi^2}\right),
\end{equation}
where $\phi=Nv/V_{g}$ is the volume fraction of the monomers, $v=b^3$ is the effective monomer volume, $\chi$ is the Flory-Huggins parameter.
In contrast to our previous works \cite{Budkov5,Budkov6}, in this study in order to take into account the volume interactions, we do not use the virial
equation of state. Instead, we use the FH interpolation formula to consider the polymer chain conformational behavior in a wide range of the monomer
volume fraction. In fact, to take into account the volume interactions between species, we might use any other interpolation formulas, such
as the Van der Waals equation of state (for example, see \cite{Budkov3}) or the virial equation of state \cite{Budkov5,Budkov6}. On the other hand, we chose the FH formula because that it allows us to relate easily the solvent quality with the parameters of volume interactions between the species. Remind that parameter $\chi$ determines
the effect of the Van der Waals interactions between the species, excluding the interaction between the permanent dipoles, which we will consider
explicitly (see below). It should be noted that the concept of 'disconnected' segments which was first introduced by I.M. Lifshitz
(see original work \cite{Lifshitz} and review \cite{Lifshitz1}) allows us to construct the total free energy by using different terms attributed to the conformational entropy of the polymer chain (eq. (\ref{eq:conf})) and volume interactions between the monomers (eq. (\ref{eq:Fvol})).

In this case, the electrostatic contribution to the free energy which is related to the thermal fluctuations of the permanent dipoles on the polymer backbone immersed to the dielectric background with permittivity $\varepsilon$ can be calculated for the large enough gyration volume at the level of the random phase approximation (RPA) (see Appendix and \cite{Dean_2012,Budkov6}):
\begin{equation}
\label{eq:Fel}
F_{el}(R_{g})\simeq\frac{2\pi k_{B}T V_{g}}{3v}\ln\left(1+\frac{4\pi p^2 \phi}{3k_{B}Tv\varepsilon}\right),
\end{equation}
where $\varepsilon$ is the reference dielectric permittivity of dielectric medium within the polymer volume which in general case must depend on the volume fraction of monomers. In the present study, to take into account this dependence, we use one of the most simple interpolation formulas which was widely used in theoretical treatment of polyelectrolyte solutions \cite{Khohlov1994,Khohlov1996,Kramarenko2000,Moldakarimov2001,Kramarenko2002}:
\begin{equation}
\label{eq:eps}
\varepsilon=\varepsilon_{s}+\left(\varepsilon_{p}-\varepsilon_{s}\right)\phi,
\end{equation}
where $\varepsilon_{p}$ is the reference permittivity of the polymeric material which is not related to the orientation fluctuations of the permanent dipoles, $\varepsilon_{s}$ is the dielectric permittivity of solvent.  Thus, unlike our previous works \cite{Budkov5,Budkov6}, where we stated that dielectric permittivity in the internal polymer volume is the same as in the bulk, in the present study we take into account the dependence of dielectric permittivity in the gyration volume on the solvent volume fraction.

Therefore, minimizing the total free energy (\ref{eq:Ftot}) with respect to the radius of gyration $R_{g}$, after some algebra we arrive at the following equation
\begin{equation}
\label{eq:alpha}
\alpha^5-\alpha=\frac{2\pi\sqrt{6}}{81}N^{3/2}\alpha^6\left(-\ln{(1-\phi)}-\phi-\chi \phi^2\right)\nonumber
\end{equation}
\begin{equation}
-\frac{4\sqrt{6}\pi^2}{243}N^{3/2}\alpha^6\left(\ln\left(1+\frac{4\pi p^2 \phi}{3k_{B}Tv\varepsilon}\right)-
\frac{\frac{4\pi p^2 \phi}{3k_{B}Tv\varepsilon}}{1+\frac{4\pi p^2 \phi}{3k_{B}Tv\varepsilon}}\frac{\varepsilon_{s}}{\varepsilon}\right).
\end{equation}

The first term in the right hand side of eq. (\ref{eq:alpha}) determines the influence of volume interactions (excluded volume and Van der Waals interactions) on the polymer chain conformation. The second term is related to the many-body electrostatic dipole correlations of monomers.

\section{Analysis of expanded coil regime}
Before we proceed to the numerical analysis of eq. (\ref{eq:alpha}), it is interesting to discuss the regime of the expanded coil conformation, i.e., when the expansion factor is $\alpha\gg 1$. In this case we obtain the following equation
\begin{equation}
\label{eq:alpha_coil}
\alpha^5-\alpha=\frac{3\sqrt{6}}{2\pi}\sqrt{N}\left(1-2\chi-\frac{32\pi^3p^4}{27(k_{B}T)^2\varepsilon_{s}^2v^2}+\frac{16\pi^2p^2}{9k_{B}T\varepsilon_{s}v}\delta\right),
\end{equation}
where the {\sl dielectric mismatch} parameter $\delta =\left(\varepsilon_{s}-\varepsilon_{p}\right)/\varepsilon_{s}$ is introduced.
Thus, for the coil conformation we can introduce the second virial coefficient of the monomer-monomer interactions as follows:
\begin{equation}
\label{eq:Beff}
B=v\left(\frac{1}{2}-\chi\right)-\frac{16\pi^3p^4}{27(k_{B}T)^2\varepsilon_{s}^2v}+\frac{8\pi^2p^2}{9k_{B}T\varepsilon_{s}}\delta.
\end{equation}
The first term in the right hand side of (\ref{eq:Beff}) is a second virial coefficient within the FH theory \cite{Khokhlov_book}. The second term is a contribution of the Keesom dipole-dipole interaction \cite{Stockmayer} which is always a negative value \cite{Schiessel1998,Dua2014,Budkov6}. The third term is related to the dielectric mismatch between the pure solvent and the polymeric material. The sign of the latter contribution is determined by the sign of the mismatch parameter $\delta$. Thus, the dielectric mismatch effect competes with the Keesom attraction between the monomers, when the condition $\varepsilon_{s}>\varepsilon_{p}$ ($\delta >0$) is satisfied. In other words, the dielectric mismatch effect tends to expand the polymer coil, whereas the Keesom monomer-monomer attraction, oppositely, provokes its shrinking. The latter can be interpreted as follows. In case of $\delta >0$ the electric charge of the permanent dipoles is preferentially solvated by the solvent molecules that leads to effective repulsive interaction between monomers. In the opposite case of $\varepsilon_{s}<\varepsilon_{p}$ ($\delta <0$), both the Keesom interaction and the dielectric mismatch effect lead to a polymer coil collapse. In this case the electric charge of the permanent dipoles tends to be solvated by the monomers instead of the solvent molecules. Thus, such polymer self-solvation leads to an additional effective attractive interaction between the monomers. This is an example of manifestation of the so-called solvation forces \cite{Israelachvili}.
It should be noted that only at $\delta=\left(\varepsilon_{s}-\varepsilon_{p}\right)/\varepsilon_{s} \leq (\varepsilon_{s}-1)/\varepsilon_{s}$ (where $\varepsilon_{s}\geq 1$) the physical condition \cite{Landau_VIII} for the dielectric permittivity of polymeric material $\varepsilon_{p}\geq1$ is fulfilled.

In conclusion of this section it is instructive to write the expression for electrostatic free energy in the limit of $\alpha\gg 1$ (or $\phi \ll 1$):
\begin{equation}
\label{eq:Fel_coil}
F_{el}\simeq \frac{8\pi^2 Np^2}{9\varepsilon_{s}v}\left(\delta -\frac{2\pi p^2}{3k_{B}T\varepsilon_{s}v}\right)\phi,
\end{equation}
where we have omitted the term which does not depend on the gyration radius. The latter relation determines electrostatic free energy of polymer coil at the level of pairwise correlations.

Therefore, one can conclude that dipole electrostatic correlations of monomers can be taken into account as pairwise ones only when the polymer chain is in the coil conformation.

\section{Numerical results and discussions}
Turning to a numerical analysis of (\ref{eq:alpha}), we define the following 'coupling' parameter
$$\lambda =\frac{p^2}{3 \varepsilon_{s}vk_{B}T}$$
which determines the 'strength' of the dipole electrostatic correlations of monomers. It should be noted that for the weak polyelectrolytes in the regime of counterion condensation, the coupling parameter $\lambda$ can be related to the Bjerrum length $l_{B}=e^2/(\varepsilon_{s}k_{B}T)$. Indeed, if we assume that $p\sim |z| e b$ ($e$ is the elementary charge, $z$ is the counterion (monomer) valency), then we arrive at $\lambda \sim z^2 l_{B}/(3 b)$. Thus, in this case the coupling parameter is proportional to the Coulomb strength mentioned in Introduction.

In this study we will consider only the case of good solvent ($\chi\leq 1/2$). Fig. 2 demonstrates the dependences of the expansion factor $\alpha$ on the coupling parameter $\lambda$ at different values of dielectric mismatch parameter $\delta$.  In the region of $\delta >0$ ($\varepsilon_{s}>\varepsilon_{p}$) we can see a pronounced maximum on the expansion factor curve. This maximum is caused by the above-mentioned competition of the Keesom dipole-dipole attraction and effective repulsion between the monomers, which is attributed to the difference between the bulk and local dielectric permittivities (see the discussion in Section 3). For negative $\delta$, the expansion factor monotonically decreases at the coupling parameter increase. In both cases, if the coupling parameter is large enough, the CG transition goes continuously. As is seen from Fig. 2, the coupling parameter value corresponding to the polymer chain collapse is very sensitive to the value of dielectric mismatch parameter $\delta$.

\begin{figure}
\centerline{\includegraphics[scale = 0.5]{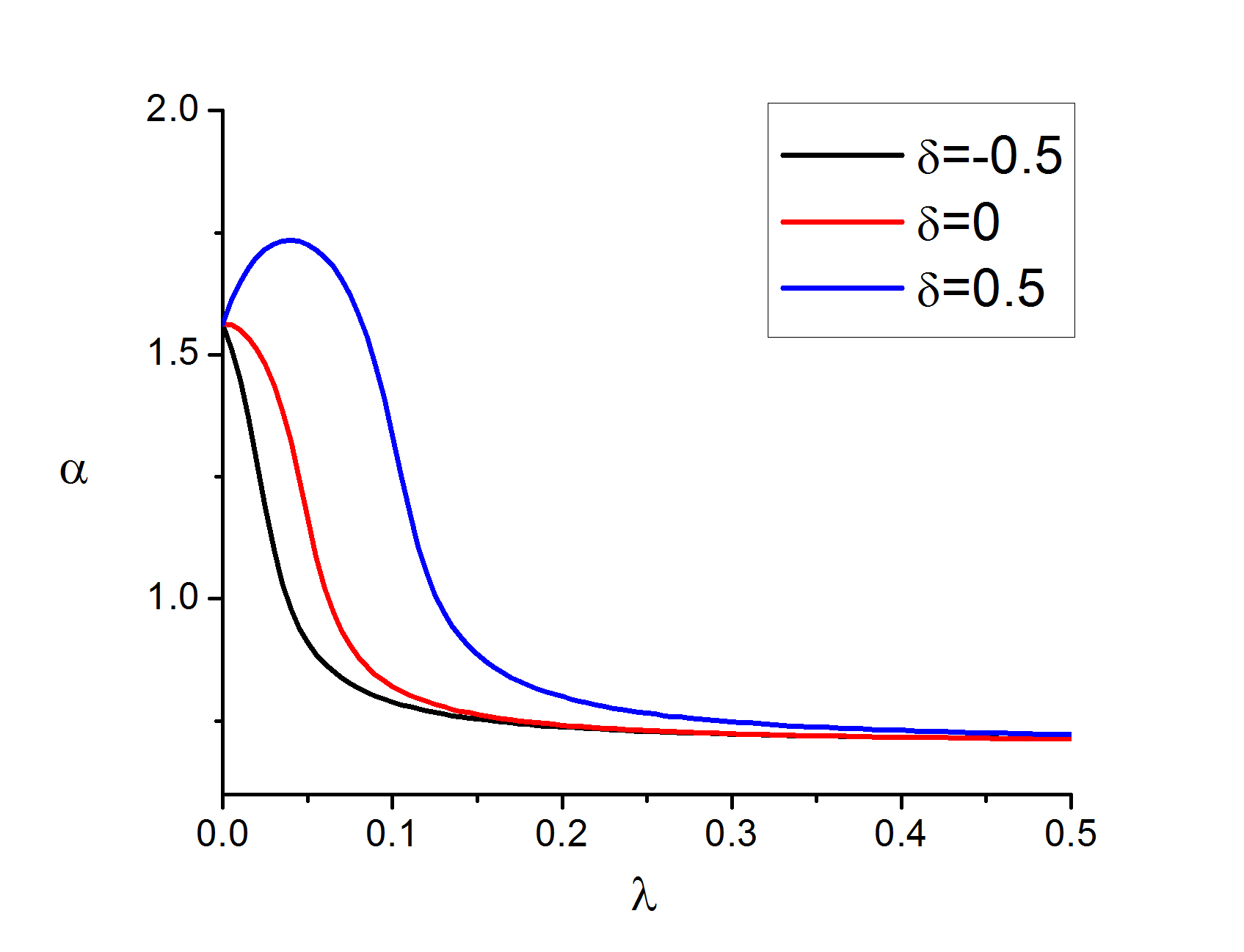}}
\caption{Dependences of expansion factor $\alpha$ on the coupling parameter $\lambda = p^2/(3 \varepsilon_{s}vk_{B}T)$ at different dielectric mismatch parameter values $\delta =\left(\varepsilon_{s}-\varepsilon_{p}\right)/\varepsilon_{s}$. The coupling parameter value at which the CG transition takes place is very sensitive to the dielectric mismatch parameter. The data are shown for $N=100$, $\chi=0.2$.}
\label{fig.2}
\end{figure}

Now we would like to compare the results obtained by the present theory with those predicted by the theory with pairwise dipole correlations (see, for instance, \cite{Schiessel1998,Dua2014}). It should be noted, that within such theory the electrostatic contribution to the free energy may be assessed by relation (\ref{eq:Fel_coil}).  As it is shown in Fig. 3, accounting for the dipole correlations at the many-body level produces a qualitatively different dependence of the expansion factor on the coupling parameter. Indeed, accounting for the dipole correlations at the pairwise level results in an abrupt decrease in the expansion factor, when the coupling parameter exceeds some threshold value. However, accounting for the higher dipole correlations makes the CG transition smoother.

\begin{figure}
\centerline{\includegraphics[scale = 0.5]{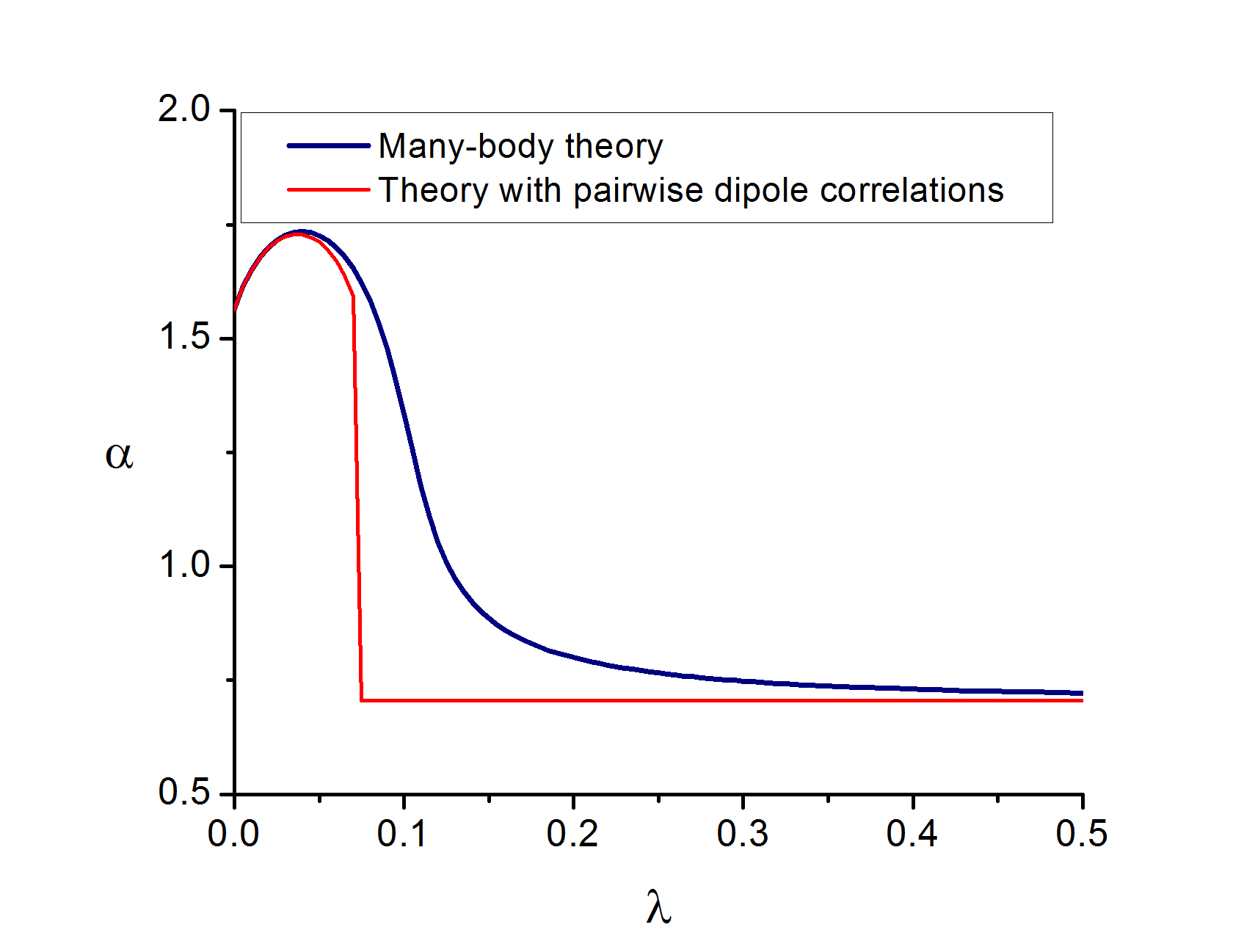}}
\caption{Dependences of expansion factor $\alpha$ on the coupling parameter $\lambda = p^2/(3 \varepsilon_{s}vk_{B}T)$ calculated within the present theory and the theory with pairwise dipole correlations of monomers. Accounting for the dipole correlations at the many-body level makes the coil-globule transition induced by dipole correlations smoother. The data are shown for $\delta=0.5$, $\chi=0.2$ and $N=100$.}
\label{fig.3}
\end{figure}

\section{Concluding remarks}
In conclusion, we have formulated a simple analytical self-consistent field theory of flexible polymer chain dissolved in a good solvent, whose monomers carry permanent freely oriented dipoles. Such effects as dipole correlations at the many-body level and effect of the dielectric heterogeneity near the polymer backbone ({\sl dielectric mismatch} effect) have been taken into account. We have shown that in the regime of good solvent strong enough electrostatic dipole correlations of monomers lead to coil-globule transition. However, in contrast to the theory with pairwise dipole correlations, predicting the coil-globule transition as a first-order phase transition \cite{Dua2014} (which becomes true phase transition only at
$N\rightarrow \infty$ \cite{Khokhlov_book}), the present many-body level theory describes the conformational transition as a continuous process. We have demonstrated that the dipole correlations of monomers can be taken into account as pairwise ones only when the polymer chain is in the coil conformation. For the globular conformation, dipole correlations must be taken into account at the many-body level. We have shown that in the case, when the solvent dielectric permittivity larger than the permittivity of the polymeric material, the dielectric mismatch effect competes with the electrostatic dipole correlations of monomers, expanding the polymer coil. In the opposite case, when the polymeric dielectric permittivity exceeds the solvent dielectric permittivity, both the dielectric mismatch effect and the dipole correlations lead to polymer chain collapse. We have found out that accounting for the dipole correlations at the many-body level smooths this pure electrostatics driven coil-globule transition of the polymer chain. We have shown that the value of the coupling parameter corresponding to the polymer chain collapse is highly sensitive to the dielectric mismatch parameter value. In our opinion, it is a very important fundamental result which may be of use for deeper understanding of the conformational behavior of biological macromolecules in the aqueous media.

It should be noted that in the present study in contrast to papers \cite{Kumar_2014,Mahalik2016} (where the dielectric permittivities were calculated within the mean-field theory), we introduce the reference dielectric permittivities for the solvent and polymeric material as the free model parameters. The latter is due to the fact that all the existing 'first-principle' analytical approaches to dielectric permittivity calculation cannot give a satisfactory agreement with the experimental values without introducing some additional fitting parameters \cite{Barrat_Hansen,Andelman_2007,Andelman_2012}. Moreover, the static dielectric permittivities of real liquid-phase solutions are determined by many different physical effects, including electronic polarizability, thermal orientational fluctuations of permanent dipoles, hydrogen bonding, etc. It is evident that all of these effects cannot be properly taken into account simultaneously within one analytical theory. Therefore, in our opinion, for the phenomenological description of thermodynamic and conformational properties of polyelectrolytes and polar polymers dissolved in some dielectric solvent it is most natural to introduce the dielectric permittivities of the solvent and polymeric material as the free model parameters.

In order to avoid the ultraviolet divergence in the electrostatic free energy within the RPA, we have introduced an ultraviolet cut-off parameter $\Lambda=2\pi/b$ which is inversely proportional to the monomer length scale (see Appendix). The choice of the cut-off parameter value is motivated by the fact that at the scales $\sim b$ there are no fluctuations of the electrostatic potential related to the thermal orientation fluctuations of dipoles \cite{Dean_2012}. The latter is quite a reasonable procedure based on a physical assumption. Nevertheless, there is another way to escape from the ultraviolet divergences in such kind of theories, leading to qualitatively the same results. Namely, instead of the point charges of the dipoles we can consider the smeared charges with some form-factors $\Gamma(\bold{k})$ \cite{Wang,Delaney2016}. Such form-factors must be quickly approaching to zero in the region of $|\bold{k}|\geq 1/b$. The latter provides the electrostatic free energy convergence in the ultraviolet limit. Undoubtedly, the choice of such form-factor is related to some arbitrariness. However, in our point of view the necessity to introduce the charge form-factors  to the theory is motivated by the fact that at the small scales (order of particle size) we cannot describe the dipole interactions (as well as the interactions related to the higher multipoles) within the classical physics framework. That is why to obtain a finite value of the electrostatic free energy, we have to phenomenologically introduce some charge form-factor, having a pure quantum nature.

In the present study we have considered the polymer chain collapse only in the case of good solvent. Nevertheless, it is interesting to discuss what conformational behavior would be take observed, when the solvent becomes poor. However, the details of this investigation deserve a separate detailed publication and, thereby, will be published elsewhere.

Finally, we would like to discuss the possible applications of the present theoretical model. Firstly, it can be used as a theoretical background for analysis of conformational behavior of weak polyelectrolyte chains in the regime of counterion condensation or polyzwitterionic macromolecules in the dilute solutions. Secondly, this theory might be combined with counterion-fluctuation theory \cite{Brilliantov1998,Brilliantov2016}. Moreover, it is interesting to find and study a 'crossover' region between the Coulombic and dipole regimes of the polyelectrolyte chain coil-globule transition. The latter is the subject of forthcoming publications.

\section{Appendix: Derivation of electrostatic free energy (\ref{eq:Fel})}
Here we present a derivation of electrostatic contribution (\ref{eq:Fel}) to the solvation free energy of the polymer chain, which is related to the monomer dipole correlations. We would like to stress that the theory which will be presented below can be applied to both the single very long polar polymer chain and the concentrated solution of overlapped polar polymer chains.

The electrostatic contribution can be expressed via the electrostatic partition function
\begin{equation}
\label{eq:Qel}
Q_{el}=\left<\exp\left[-\frac{1}{2k_{B}T}\int_{V}d\bold{r}\int_{V}d\bold{r}^{\prime}\hat{\rho}(\bold{r})G_{0}(\bold{r},\bold{r}^{\prime})\hat{\rho}(\bold{r}^{\prime})\right]\right>,
\end{equation}
as follows
\begin{equation}
\label{eq:Fel2}
F_{el}=-k_{B}T\ln Q_{el},
\end{equation}
where the operator-inverse $G_{0}(\bold{r},\bold{r}^{\prime})$ with respect to the operator
\begin{equation}
\label{eq:G0^-1}
G_{0}^{-1}(\bold{r},\bold{r}^{\prime})=-\frac{1}{4\pi}\nabla\left(\varepsilon(\bold{r})\nabla\delta(\bold{r}-\bold{r}^{\prime})\right)
\end{equation}
and the microscopic dipole charge density
\begin{equation}
\label{eq:rho_dip}
\hat{\rho}(\bold{r})=-\sum\limits_{j=1}^{N}\bold{p}_{j}\nabla\delta\left(\bold{r}-\bold{r}_{j}\right)
\end{equation}
are introduced; $\varepsilon(\bold{r})$ is the medium reference dielectric permittivity, which is not related to the permanent dipoles of the monomers; symbol $\left<..\right>$ means the average over the orientations of noninteracting permanent dipoles and the positions of monomers.

Using the standard Hubbard-Stratonovich transformation, we rewrite the electrostatic partition function as the following functional integral
\begin{equation}
\label{eq:HS}
Q_{el}=\int\frac{\mathcal{D}\psi}{C}\exp\left[-\frac{k_{B}T}{2}\left(\psi, G_{0}^{-1}\psi\right)\right]\left<\exp\left[i\left(\hat{\rho},\psi\right)\right]\right>,
\end{equation}
where the short-hand notations
\begin{equation}
\left(\psi, G_{0}^{-1}\psi\right)=\int_{V}d\bold{r}\int_{V}d\bold{r}^{\prime} \psi(\bold{r})G_{0}^{-1}(\bold{r},\bold{r}^{\prime})\psi(\bold{r}^{\prime}),
\end{equation}
and
\begin{equation}
\left(\hat{\rho},\psi\right)=\int_{V}d\bold{r}\hat{\rho}(\bold{r})\psi(\bold{r})
\end{equation}
are introduced; $C=\int\mathcal{D}\psi\exp\left[-\frac{k_{B}T}{2}\left(\psi, G_{0}^{-1}\psi\right)\right]$ is the normalization constant.

Applying the standard cumulant expansion in integrand of (\ref{eq:HS}) and truncating it at second order, we obtain
\begin{equation}
\nonumber
\left<\exp\left[i\left(\hat{\rho},\psi\right)\right]\right>=\exp\left[i\int_{V}d\bold{r}\left<\hat{\rho}(\bold{r})\right>_{c}\psi(\bold{r})-\frac{1}{2}\int_{V}d\bold{r}\int_{V}d\bold{r}^{\prime} \left<\hat{\rho}(\bold{r})\hat{\rho}(\bold{r}^{\prime})\right>_{c}\psi(\bold{r})\psi(\bold{r}^{\prime})+..\right]
\end{equation}
\begin{equation}
\nonumber
=\exp\left[-\frac{1}{2}\int_{V}d\bold{r}\int_{V}d\bold{r}^{\prime} \sum\limits_{i=1}^{N}\sum\limits_{j=1}^{N}\left<\bold{p}_{i}^{\alpha}\bold{p}_{j}^{\gamma}\right>\left<\nabla_{\alpha}\delta(\bold{r}-\bold{r}_{i})\nabla_{\gamma}\delta(\bold{r}^{\prime}-\bold{r}_{j})\right>\psi(\bold{r})\psi(\bold{r}^{\prime})+..\right]
\end{equation}
\begin{equation}
\label{eq:cumul}
=\exp\left[-\frac{p^2}{6}\sum\limits_{j=1}^{N}\left<\left(\nabla\psi(\bold{r}_{j})\right)^2\right>+..\right]=\exp\left[-\frac{p^2}{6}\int_{V}d\bold{r}\left<\hat{n}_{m}(\bold{r})\right>\left(\nabla\psi(\bold{r})\right)^2+..\right],
\end{equation}
where the relations $\left<\hat{\rho}(\bold{r})\right>=0$ and $\left<\bold{p}_{i}^{\alpha}\bold{p}_{j}^{\gamma}\right>=p^2\delta_{\alpha\gamma}\delta_{ij}/3$ ($\alpha,\gamma=1,2,3$ and $i,j=1,..,N$) have been taken into account and the microscopic monomer density $\hat{n}_{m}(\bold{r})=\sum_{j=1}^{N}\delta(\bold{r}-\bold{r}_{j})$ has been introduced; symbol $\left<..\right>_{c}$ means the cumulant average \cite{Kubo}; $\delta_{\alpha\gamma}$ and $\delta_{ij}$ are the Kronecker delta.

Therefore, at the level of Gaussian approximation we obtain
\begin{equation}
\nonumber
Q_{el}\approx\int\frac{\mathcal{D}\psi}{C}\exp\left[-\frac{k_{B}T}{2}\left(\psi, G_{0}^{-1}\psi\right)\right]\exp\left[-\frac{p^2}{6}\int_{V}d\bold{r}\left<\hat{n}_{m}(\bold{r})\right>\left(\nabla\psi(\bold{r})\right)^2\right]
\end{equation}
\begin{equation}
=\int\frac{\mathcal{D}\psi}{C}\exp\left[-\frac{k_{B}T}{2}\left(\psi, G^{-1}\psi\right)\right]=\sqrt{\frac{\det{G_{0}^{-1}}}{\det{G^{-1}}}},
\end{equation}
where the operator
\begin{equation}
\label{eq:G^-1}
G^{-1}(\bold{r},\bold{r}^{\prime})=-\frac{1}{4\pi}\nabla\left(\varepsilon_{r}(\bold{r})\nabla\delta(\bold{r}-\bold{r}^{\prime})\right),
\end{equation}
and renormalized dielectric permittivity \cite{Budkov2015,Budkov2016_2}
\begin{equation}
\label{eq:eps_ren}
\varepsilon_{r}(\bold{r})=\varepsilon(\bold{r})+\frac{4\pi p^2}{3k_{B}T}\left<\hat{n}_{m}(\bold{r})\right>
\end{equation}
have been introduced.

In the case of large enough system volume ($V\rightarrow \infty$) and homogeneous dielectric medium ($\varepsilon(\bold{r})=\varepsilon=const$) neglecting the boundary effects, we get the following estimate for the electrostatic free energy
\begin{equation}
\label{eq:Fel2}
F_{el}\simeq -\frac{Vk_{B}T}{2}\int\limits_{|\bold{k}|<\Lambda}\frac{d\bold{k}}{(2\pi)^3}\ln\left(\frac{G(\bold{k})}{G_{0}(\bold{k})}\right),
\end{equation}
where $G_{0}(\bold{k})=4\pi/(\varepsilon\bold{k}^2)$ and $G(\bold{k})=4\pi/(\varepsilon_{r}\bold{k}^2)$ are the Fourier-images of the Green functions of Poisson equation for the infinite space; $\varepsilon_{r}=\varepsilon+4\pi p^2 n_{m}/(3k_{B}T)$ is the renormalized dielectric permittivity; $n_{m}=\left<\hat{n}_{m}(\bold{r})\right>=N/V$ is the average monomer number density; $\Lambda=2\pi/b$ is the parameter of ultraviolet cut-off. The choice of such value of the cut-off parameter $\Lambda$ is motivated by the fact that at the scales $\sim b$ there are no fluctuations of the electrostatic potential related to the thermal orientation fluctuations of dipoles \cite{Dean_2012}. Using the above expressions, we eventually obtain
\begin{equation}
\label{eq:Fel_final}
F_{el}\simeq \frac{2\pi Vk_{B}T}{3b^3}\ln\left(1+\frac{4\pi p^2 n_{m}}{3k_{B}T\varepsilon}\right).
\end{equation}
Expression (\ref{eq:Fel_final}) determines the electrostatic contribution to the total free energy of the polymer solution related to the dipole correlations of the monomers at the level of Gaussian approximation (RPA).

\begin{acknowledgments}
YAB thanks N.V. Brilliantov for insightful discussions. YAB designed the study, participated in the derivation of the main analytical results and
drafted the manuscript. NNK participated in the derivation of the analytical results, carried out numerical calculations and helped to draft the manuscript. ALK participated in the discussions, derivation of the analytical results and helped to draft the manuscript. This work was supported by grant from Russian Foundation for Basic Research (Grant No. 15-43-03195).
\end{acknowledgments}

\newpage


\begin{thebibliography}{99}
\bibitem{Tomasko}
{D.L. Tomasko, H. Li, et al.}, Ind. Eng. Chem. $\bold{42}$ (2003) 6431.

\bibitem{Kazarian}
{Kazarian S. G.}, Polymer Science, Ser C, $\bold{42}$ 1 (2000) 78.

\bibitem{Kiran2016}
Erdogan Kiran J. of Supercritical Fluids $\bold{110}$ (2016) 126.

\bibitem{DrugDeliveryReview1}
{Kost J., Langer R.} Advanced Drug Delivery Reviews $\bold{46}$ (2001) 125.

\bibitem{DrugDeliveryReview2}
{Priya Bawa, Viness Pillay, Yahya E Choonara and Lisa C du Toit}  Biomed. Mater. $\bold{4}$ (2009) 022001.

\bibitem{DrugDeliveryReview3}
{ Fitzpatrick S.D., Lindsay E.F., Thakur A., et.al.}  Expert Review of Medical Devices  $\bold{9}$ (4) (2012) 339.


\bibitem{deGennes_collaps}
{de Gennes P.G.} Le Journal De Physique - Letters $\bold{36}$ (2) (1975) L-55.

\bibitem{Grosberg}
{Grosberg A.Yu., Kuznetsov D.V.}  Macromolecules $\bold{25}$ (1992) 1970.

\bibitem{Birshtein}
{Birshtein T.M., Pryamitsyn V.A.}  Macromolecules $\bold{24}$ (1991) 1554.

\bibitem{Moore}
{Moore M.A.} J. Phys. A: Math. Gen. $\bold{10}$ (2) (1977) 305.

\bibitem{Sanchez}
{Sanchez I.C.} Macromolecules $\bold{12}$ (5) (1979) 980.

\bibitem{Lifshitz}
{Lifshitz I.M.} Soviet Physics JETP $\bold{28}$ (6) (1969) 1280.

\bibitem{Lifshitz1}
{Lifshitz I.M., Grosberg A.Yu., A.R. Khohlov} Rev. Mod. Phys. $\bold{50}$ 3 (1978) 683.

\bibitem{Muthukumar}
{Muthukumar M.} J. Chem. Phys. $\bold{81}$ (1984) 6272.

\bibitem{Dua}
{Dua A. and Cherayil B.J.} J. Chem. Phys. $\bold{111}$ (7) (1999) 3274.

\bibitem{Vilgis2005}
{A. Dua and T. A. Vilgis} EPL $\bold{71}$ (1) (2005) 49.

\bibitem{Simmons2013}
{Simmons D.S., Sanchez I.C.} Macromolecules $\bold{46}$ (2013) 4691.

\bibitem{Matsuyama1990}
{Akihiko Matsoyanla and Fumihiko Tanaka} J. Chem. Phys. $\bold{94}$ (1) (1991) 781.

\bibitem{Muzdalo}
{Heyda J., Muzdalo A., Dzubiella J.} Macromolecules $\bold{46}$ (2013) 1231.

\bibitem{Tamm}
{Tamm M.V., Erukhimovich I.Ya.} Polymer Science, Ser. A $\bold{44}$ (2) (2002) 196.

\bibitem{Budkov1}
{Budkov Yu.A., Kolesnikov A.L., Georgi N., and Kiselev M.G.} J. Chem. Phys. $\bold{141}$ (2014) 014902.

\bibitem{Budkov2}
{Budkov Yu.A., Vyalov I.I., Kolesnikov A.L., Georgi N., Chuev G.N., Kiselev M.G.} J. Chem. Phys. $\bold{141}$ (2014) 204904.

\bibitem{Budkov3}
{Budkov Yu.A., Kolesnikov A.L., Georgi N., Kiselev M.G.} EPL $\bold{109}$ (2015) 36005.

\bibitem{Budkov4}
{Yu. A. Budkov, A. L. Kolesnikov, N. N. Kalikin and M. G. Kiselev} EPL $\bold{114}$ (2016) 46004.

\bibitem{Melnikov1999}
{Sergey M. Mel'nikov, Malek O. Khan, Bjorn Lindman, and  Bo Jonsson} J. Am. Chem. Soc. $\bold{121}$ (1999) 1130.

\bibitem{Borkovec2004}
{Luke J. Kirwan, Georg Papastavrou, and Michal Borkovec} Nano Letters $\bold{4}$ (1) (2004) 149.

\bibitem{Brilliantov2016}
{Anvy Moly Tom, Satyavani Vemparala, R. Rajesh, Nikolai V. Brilliantov} Phys. Rev. Lett. $\bold{117}$ (7) (2016) 147801.

\bibitem{Gavrilov2016}
{A. A. Gavrilov, A. V. Chertovich, and E. Yu. Kramarenko} Macromolecules $\bold{49}$ (3) (2016) 1103.

\bibitem{Netz_2003}
{Netz R.R.} J. Phys. Chem. B $\bold{107}$ (2003) 8208.

\bibitem{Netz_2003_2}
{R. R. Netz} Phys. Rev. Lett. $\bold{90}$ (2003) 128104.

\bibitem{Manning1978}
{G. Manning} Q. Rev. Biophys. $\bold{11}$ (1978) 179.

\bibitem{Brilliantov1998}
{Brilliantov N.V., Kuznetzov D.V., Klein R.} Phys. Rev. Lett. $\bold{81}$ (7) (1998) 1433.

\bibitem{Levin2002}
{Y. Levin} Rep. Prog. Phys. $\bold{65}$ (2002) 1577.

\bibitem{Brilliantov1993}
{N.V. Brilliantov} Phys. Rev. E $\bold{48}$ (6) (1993) 4536.

\bibitem{Liverpool1999}
{Ramin Golestanian, Mehran Kardar, and Tanniemola B. Liverpool} Phys. Rev. Lett. $\bold{82}$ (22) (1999) 4456.

\bibitem{Kirkwood1952}
{J. Kirkwood and J.B. Shumaker} Proc. Natl. Acad. Sci. USA $\bold{38}$ (1952) 855.

\bibitem{Adzic2014}
{Natasa Adzic and Rudolf Podgornik} Eur. Phys. J. E $\bold{37}$ (2014) 49.

\bibitem{Budkov_colloid}
{Yu. A. Budkov, A. I. Frolov, M. G. Kiselev, and N. V. Brilliantov} J. Chem. Phys. $\bold{139}$ (2013) 194901.

\bibitem{Budkov_polyel}
{Yu. A. Budkov, A. L. Kolesnikov, N. Georgi, E. A. Nogovitsyn, and M. G. Kiselev} J. Chem. Phys. $\bold{142}$ (2015) 174901.

\bibitem{Zhang2016}
{Pengfei Zhang, Nayef M. Alsaifi, Jianzhong Wu, and Zhen-Gang Wang} Macromolecules  $\bold{49}$ (24) (2016) 9720.

\bibitem{Shen2017}
{Kevin Shen and Zhen-Gang Wang} J. Chem. Phys. $\bold{146}$ (2017) 084901.

\bibitem{Muthu_collapse}
{Arindam Kundagrami and M. Muthukumar} Macromolecules $\bold{43}$ (2010) 2574.

\bibitem{BrilliantovOCP}
{N. V. Brilliantov} Contrib. Plasma Phys. $\bold{38}$(4) (1998) 489.

\bibitem{Brilliantov2017}
{Anvy Moly Tom, Satyavani Vemparala, R. Rajeshab and Nikolai V. Brilliantov} Soft Matter $\bold{13}$ (2017) 1862.

\bibitem{Schiessel1998}
{H. Schiessel and P. Pincus} Macromolecules $\bold{31}$ (22) (1998) 7953.

\bibitem{Dua2014}
{Prasanta Kundu, Arti Dua} J. Stat. Mech. $\bold{2014}$ (2014) 07023.

\bibitem{Cherstvy2010}
{A. G. Cherstvy} J. Phys. Chem. B $\bold{114}$ (16) (2010) 5241.

\bibitem{Kumar_2009}
{Rajeev Kumar and Glenn H. Fredrickson} J. Chem. Phys. $\bold{131}$ (2009) 104901.

\bibitem{Budkov6}
{Yu.A. Budkov and A.L. Kolesnikov} Eur. Phys. J. E $\bold{39}$ (2016) 110.

\bibitem{Podgornik_2004}
{Rudi Podgornik} Phys. Rev. E $\bold{70}$ (2004) 031801.

\bibitem{Dean_2012}
{David S. Dean and Rudolf Podgornik} J. Chem. Phys. $\bold{136}$ (2012) 154905.

\bibitem{Kumar_2014}
{Rajeev Kumar, Bobby G. Sumpter, and M. Muthukumar} Macromolecules $\bold{47}$ (2014) 6491.

\bibitem{Lu2015}
{Bing-Sui Lu, Ali Naji, and Rudolf Podgornik} J. Chem. Phys. $\bold{142}$ (2015) 214904.

\bibitem{Delaney2016}
{Jonathan M. Martin, Wei Li, Kris T. Delaney, and Glenn H. Fredrickson} J. Chem. Phys. $\bold{145}$ (2016) 154104.

\bibitem{Mahalik2016}
{Jyoti P. Mahalik, Bobby G. Sumpter, and Rajeev Kumar} Macromolecules $\bold{49}$ (18) (2016) 7096.

\bibitem{Gurovich1}
{E. Gurovich} Macromolecules $\bold{27}$ (1994) 7339.

\bibitem{Gurovich2}
{E. Gurovich} Macromolecules $\bold{28}$ (1994) 6078.

\bibitem{Budkov5}
{Yu. A. Budkov, A. L. Kolesnikov, and M. G. Kiselev} J. Chem. Phys. $\bold{143}$ (2015) 201102.

\bibitem{Flory_book}
{Flory P.} {\sl Statistical Mechanics of Chain Molecules}  New York: Wiley-Interscience, 1969.

\bibitem{Fixman}                                                           {Fixman M.} J. Chem. Phys. $\bold{36}$ (2) 306 (1962).

\bibitem{Budkov2016}
{Budkov Yu.A., Kolesnikov A.L.} J. Stat. Mech. $\bold{2016}$ 103211 (2016).

\bibitem{Khokhlov_book}
{A. Yu. Grosberg and A. R. Khokhlov} {\sl Statistical Physics of Macromolecules} AIP
Press, Woodbury, NY, 1994.

\bibitem{Khohlov1994}
{Alexei R. Khokhlov, Elena Yu. Kramarenko} Macromol. Theory Simul. $\bold{3}$ (1994) 45.

\bibitem{Khohlov1996}
{Alexei R. Khokhlov and Elena Yu. Kramarenko} Macromolecules $\bold{29}$ (1996) 681.

\bibitem{Kramarenko2000}
{Elena Yu. Kramarenko, Alexei R. Khokhlov, Kenichi Yoshikawa} Macromol. Theory Simul. $\bold{9}$ (2000) 249.

\bibitem{Moldakarimov2001}
{Samat B. Moldakarimov, Elena Yu. Kramarenko, Alexei R. Khokhlov, Sarkyt E. Kudaibergenov} Macromol. Theory Simul. 10 (2001) 780.

\bibitem{Kramarenko2002}
{E. Yu. Kramarenko, I. Ya. Erukhimovich, A. R. Khokhlov} Macromol. Theory Simul. 11 (2002) 462.

\bibitem{Stockmayer}
{W. H. Stockmayer} J. Chem. Phys.  $\bold{9}$, (1941) 398.

\bibitem{Israelachvili}
{Jacob N. Israelachvili} {\sl Intermolecular and surface forces} (Academic Press, 2011).

\bibitem{Landau_VIII}
{Landau L.D., Lifshitz E.M.}  {\sl Electrodynamics of Continuous Media V. 8, A Course
of Theoretical Physics} (Pergamon Press, Oxford, UK, 1960).

\bibitem{Barrat_Hansen}
{Barrat J.-L., Hansen J.-P.} University Press, Cambridge: 2003.

\bibitem{Andelman_2007}
{Abrashkin A., Andelman D., Orland H.} PRL $\bold{99}$ (2007) 077801.

\bibitem{Andelman_2012}
{Levy A., Andelman D., Orland H.} PRL $\bold{108}$ (2012) 227801.

\bibitem{Wang}
{Z.-G. Wang} Phys. Rev. E $\bold{81}$ (2010) 021501.

\bibitem{Kubo}
{Kubo R.} J. Phys. Soc. Jap. $\bold{17}$ (7) (1962) 1100.

\bibitem{Budkov2015}
{Budkov Yu.A., Kolesnikov A.L., Kiselev M.G.} EPL $\bold{111}$ (2015) 28002.

\bibitem{Budkov2016_2}
{Budkov Yu.A., Kolesnikov A.L., Kiselev M.G.} J. Chem. Phys. $\bold{144}$ (2016) 184703.



\end{thebibliography}
\end{document}